\documentclass[aps,pra,superscriptaddress,floatfix,showpacs,10pt,two
column]{revtex4}%
\usepackage[dvips]{graphicx}
\usepackage{amssymb}
\usepackage{amsmath}
\usepackage{color}
\usepackage{amsfonts}%
\def\func#1{\mathop{\rm #1}\nolimits}%
\def\dsum{\mathop{\displaystyle \sum }}
\definecolor{ueblue}{rgb}{0,0,0.2}
\definecolor{hh}{rgb}{1,0.4,0.3}
\begin{document}
\title{Bounds on the multipartite entanglement of superpositions}
\author{Wei Song}
\affiliation{Hefei National Laboratory for Physical Sciences at
Microscale and Department of Modern Physics, University of Science
and Technology of China, Hefei, Anhui 230026, China}
\author{Nai-Le Liu}
\affiliation{Hefei National Laboratory for Physical Sciences at
Microscale and Department of Modern Physics, University of Science
and Technology of China, Hefei, Anhui 230026, China}
\author{Zeng-Bing Chen}
\affiliation{Hefei National Laboratory for Physical Sciences at
Microscale and Department of Modern Physics, University of Science
and Technology of China, Hefei, Anhui 230026, China}

\pacs{03.67.-a, 03.67.Mn, 03.65.Ud}

\begin{abstract}
We derive the lower and upper bounds on the entanglement of a given
multipartite superposition state in terms of the entanglement of the
states being superposed. The first entanglement measure we use is
the geometric measure, and the second is the q-squashed
entanglement. These bounds allow us to estimate the amount of the
multipartite entanglement of superpositions. We also show that two
states of high fidelity to one another do not necessarily have
nearly the same q-squashed entanglement.
\end{abstract}
\maketitle

Recently, Linden \emph{et al.} \cite{Linden:2006} raised the
following problem: given a certain decomposition of a bipartite
state $|\Gamma\rangle$ as a superposition of two other states, what
is the relation between the entanglement of $|\Gamma\rangle$ and
those of the two states being superposed? They derived upper bounds
on the entanglement of $|\Gamma\rangle$ in terms of those of the two
states in the superposition, where the entanglement was quantified
by the von Neumann entropy. Subsequently, Yu \emph{et al.}
\cite{Yu:2007} and Ou \emph{et al.} \cite{Ou:2007} investigated the
same problem in terms of concurrence and negativity, respectively.
Yet their results only apply to bipartite cases. The aim of this
paper is to provide general lower and upper bounds on the
entanglement of superposition states in multipartite scenarios.
First we use the so-called geometric measure
\cite{Shimony:1995,Barnum:2001}. This measure distinguishes itself
in that it is suitable for any-partite systems with arbitrary
dimensions. Then we use the q-squashed entanglement
\cite{Christandl:2003}. Among all existing entanglement measures,
only the q-squashed entanglement has been proved to be additive both
for bipartite states and for multipartite states \cite{Yang:2007}.

We begin by briefly reviewing the definition of geometric measure.
Given a general $k$-partite pure state $|\psi\rangle$, the geometric
measure is defined as \cite{Barnum:2001}
\begin{equation}
E_{g}(|{\psi \rangle )}=1-\Lambda _{^{\max }}^{k}(|{\psi \rangle )},
\label{eq1}
\end{equation}
where $\Lambda _{^{\max }}^{k}(|{\psi \rangle )}=\sup_{|\phi \rangle
\in S_{k}}|\langle {\psi |\phi \rangle |}^{2}$ with $S_k$ being the
set of $k$-separable states. In terms of the geometric measure we
have a lower bound on the entanglement of a multipartite
superposition state, as formulated in the following theorem:

\textbf{Theorem 1:} \emph{Let $|\psi _1\rangle $ and $|\psi
_2\rangle $ be arbitrary normalized k-partite pure states. The
geometric
measure of their superposed states $|\Gamma \rangle =\dfrac{a|{\psi _{1}\rangle }+b|{\psi _{2}\rangle }}{%
\left\Vert {a|{\psi _{1}\rangle }+b}|{\psi _{2}\rangle }\right\Vert
}$ with $|a|^{2}+|b|^{2}=1$ satisfies}

\begin{equation}
\begin{array}{l}
 \left\| {a\left| {\psi _1 } \right\rangle + b\left| {\psi _2 }
\right\rangle } \right\|^2E_g \left( {\left| \Gamma \right\rangle
} \right) \ge \max \left\{ {\left| a \right|^2E_g \left( {\left|
{\psi _1 }
\right\rangle } \right)} \right. \\
 + \left| b \right|^2E_g \left( {\left| {\psi _2 } \right\rangle } \right) +
2\left[ {Re\left( {a^\ast b\left\langle {\psi _1 } \right|\left.
{\psi _2 }
\right\rangle } \right)} \right. \\
 \left. {\left. { - \left| {ab} \right|\sqrt {1 - E_g \left( {\left| {\psi
_1 } \right\rangle } \right)} \sqrt {1 - E_g \left( {\left| {\psi
_2 }
\right\rangle } \right)} } \right],0} \right\} \\
 \end{array}\label{eq2}
\end{equation}

Proof: Suppose $\left\vert \phi \right\rangle $ is the optimal
$k$-separable state for $|\Gamma\rangle $, i.e., the separable state
closest to $|\Gamma\rangle $. Then we have
\begin{align}
\Lambda _{^{\max }}^{k}\left( {\left\vert \Gamma \right\rangle }\right) & =%
\frac{1}{\left\Vert a|{\psi _{1}\rangle +b|\psi _{2}\rangle }\right\Vert ^{2}%
}\Big\{|a|^{2}|\langle {\psi _{1}|}\phi \rangle
{|}^{2}+|b|^{2}|\langle {\psi
_{2}|}\phi \rangle {|}^{2}  \notag \\
& +2\func{Re}\big[a^{\ast }b\langle {\psi _{1}|}\phi \rangle \langle
{\phi
|\psi _{2}}\rangle \big]\Big\}  \notag \\
& \leq \frac{1}{\left\Vert a|{\psi _{1}\rangle +b|\psi _{2}\rangle }%
\right\Vert ^{2}}\Big\{|a|^{2}{\Lambda _{^{\max }}^{k}(|{\psi
_{1}\rangle })}
\notag \\
& {+|b|}^{2}{\Lambda _{^{\max }}^{k}(|{\psi _{2}\rangle })}+2|ab|\sqrt{%
\Lambda _{^{\max }}^{k}{(|{\psi _{1}\rangle })}\Lambda _{^{\max }}^{k}{(|{%
\psi _{2}\rangle })}}\Big\}. \label{eq3}
\end{align}
By some simple algebraic calculation, we obtain
\begin{eqnarray}
&&\left\Vert a|\psi _{1}\rangle +b|\psi _{2}\rangle \right\Vert
^{2}E_{g}(|\Gamma \rangle )\geq |a|^{2}E_{g}(|{\psi _{1}}\rangle )  \notag \\
&&+|b|^{2}E_{g}(|{\psi _{2}}\rangle )+2\Big[\func{Re}(a^{\ast }b\langle {%
\psi _{1}|\psi _{2}\rangle })  \notag \\
&&-|ab|{\sqrt{1-E_{g}(|{\psi _{1}}\rangle )}\sqrt{1-E_{g}(|{\psi _{2}}%
\rangle )}}\,\Big],  \label{eq4}
\end{eqnarray}
where the lower bound is saturated if $|{\psi _{1}\rangle }=|{\psi
_{2}\rangle }$. Because the geometric measure must be a nonnegative
value, the proof is completed.$\hfill\blacksquare$

Below we will derive an upper bound for the geometric measure of the
superposition state $|\Gamma\rangle$. For this purpose we use a
lower bound less stringent than that in Eq. (\ref{eq4}).  As a
consequence of Eq. (\ref{eq3}) we have
\begin{align}
\Lambda _{^{\max }}^{k}\left( {\left\vert \Gamma \right\rangle
}\right) &
\leq \frac{1}{\left\Vert a|{\psi _{1}\rangle +b|\psi _{2}\rangle }%
\right\Vert ^{2}}\big\{|a|^{2}{\Lambda _{^{\max }}^{k}(|{\psi
_{1}\rangle })}
\notag \\
& {+|b|}^{2}{\Lambda _{^{\max }}^{k}(|{\psi _{2}\rangle
})}+|ab|[\Lambda _{^{\max }}^{k}{(|{\psi _{1}\rangle })+}\Lambda
_{^{\max }}^{k}{(|{\psi _{2}\rangle })]}\big\}.  \label{eq5}
\end{align}
From this inequality it follows immediately  that
\begin{eqnarray}
&&\left\Vert a|\psi _{1}\rangle +b|\psi _{2}\rangle \right\Vert
^{2}E_{g}(|\Gamma \rangle )  \notag \\
&\geq &|a(a+b)|E_{g}\left( |\psi _{1}\rangle \right)
+|b(a+b)|E_{g}\left(
|\psi _{2}\rangle \right)   \notag \\
&&+2\big[\func{Re}(a^{\ast }b\langle \psi _{1}|\psi _{2}\rangle
)-|ab|\big]. \label{eq6}
\end{eqnarray}
Having the lower bound above, we can prove the following result.

\textbf{Theorem 2.} \emph{Let} $|\psi _1\rangle$ \emph{and}
$|\psi_2\rangle$ \emph{be arbitrary normalized k-partite pure
states. The geometric measure of their
superposed states} $|\Gamma \rangle =\dfrac{a|{\psi _{1}\rangle }+b|{\psi _{2}\rangle }}{%
\left\Vert {a|{\psi _{1}\rangle }+b}|{\psi _{2}\rangle }\right\Vert
}$ \emph{with} $|a|^{2}+|b|^{2}=1$  \emph{satisfies}
\begin{equation}
\left\Vert {a}|{{\psi _{1}\rangle }+b}|{{\psi _{2}\rangle
}}\right\Vert ^{2}E_{g}(|{\Gamma \rangle )}\leq \min
\big\{{A,B,\left\Vert {a}|{{\psi _{1}\rangle }+b}|{{\psi _{2}\rangle
}}\right\Vert ^{2}\big\}},  \label{eq7}
\end{equation}
where
\begin{eqnarray}
A &=&\dfrac{1}{\left\vert \left\Vert a|{\psi _{1}\rangle +b|\psi
_{2}\rangle }\right\Vert -b\right\vert }\big\{|a|^{2}E_{g}{(|{\psi
_{1}\rangle })}
\notag \\
&&-|b|\left\vert \left\Vert a|{\psi _{1}\rangle +b|\psi _{2}\rangle }%
\right\Vert -b\right\vert E_{g}{(|{\psi _{2}\rangle })}  \notag \\
&&+2\big[\func{Re}\big(a^{\ast }b\langle {\psi _{1}|\psi
_{2}}\rangle
+|b|^{2}\big)  \notag \\
&&+|b|\left\Vert a|{\psi _{1}\rangle +b|\psi _{2}\rangle }\right\Vert \big]%
\big\}  \label{a}
\end{eqnarray}
and
\begin{eqnarray}
B &=&\dfrac{1}{\left\vert \left\Vert a|{\psi _{1}\rangle +b|\psi
_{2}\rangle }\right\Vert -a\right\vert }\big\{|b|^{2}E_{g}{(|{\psi
_{2}\rangle })}
\notag \\
&&-|a|\left\vert \left\Vert a|{\psi _{1}\rangle +b|\psi _{2}\rangle }%
\right\Vert -a\right\vert E_{g}{(|{\psi _{1}\rangle })}  \notag \\
&&+2\big[\func{Re}\big(ab^{\ast }\langle {\psi _{2}|\psi
_{1}}\rangle
+|a|^{2}\big)  \notag \\
&&+|a|\left\Vert a|{\psi _{1}\rangle +b|\psi _{2}\rangle }\right\Vert \big]%
\big\}.  \label{b}
\end{eqnarray}

Proof: To prove theorem 2, it is convenient to rewrite $|\psi
_1\rangle$ as
\begin{equation}
\left\vert {\psi _{1}}\right\rangle =\frac{\frac{\left\Vert {a\left\vert {%
\psi _{1}}\right\rangle +b\left\vert {\psi _{2}}\right\rangle }\right\Vert }{%
\sqrt{\left\Vert {a\left\vert {\psi _{1}}\right\rangle +b\left\vert
{\psi
_{2}}\right\rangle }\right\Vert ^{2}+\left\vert b\right\vert ^{2}}}}{\frac{a%
}{\sqrt{\left\Vert {a\left\vert {\psi _{1}}\right\rangle
+b\left\vert {\psi
_{2}}\right\rangle }\right\Vert ^{2}+\left\vert b\right\vert ^{2}}}}%
\left\vert \Gamma \right\rangle -\frac{\frac{b}{\sqrt{\left\Vert {%
a\left\vert {\psi _{1}}\right\rangle +b\left\vert {\psi _{2}}\right\rangle }%
\right\Vert ^{2}+\left\vert b\right\vert ^{2}}}}{\frac{a}{\sqrt{\left\Vert {%
a\left\vert {\psi _{1}}\right\rangle +b\left\vert {\psi _{2}}\right\rangle }%
\right\Vert ^{2}+\left\vert b\right\vert ^{2}}}}\left\vert {\psi _{2}}%
\right\rangle .  \label{eq8}
\end{equation}
Applying Eq. (\ref{eq6}) to $|\psi _1\rangle$, we get
\begin{eqnarray}
|a|^{2}E_{g}(|{{\psi _{1}\rangle })} &\geq &\left\Vert {a|{{\psi
_{1}\rangle
}}+b}|{{\psi _{2}\rangle }}\right\Vert   \notag \\
&&\times \left\vert \left\Vert {a|{{\psi _{1}\rangle }}+b}|{{\psi
_{2}\rangle }}\right\Vert {-b}\right\vert E_{g}(|{\Gamma \rangle )}
\notag
\\
&&+|b|\left\vert \left\Vert {a|{{\psi _{1}\rangle }}+b}|{{\psi _{2}\rangle }}%
\right\Vert {-b}\right\vert E_{g}(|{{\psi _{2}\rangle })}  \notag \\
&&-2\big[\func{Re}({{a}^{\ast }{b}\langle {{\psi _{1}}}|{{\psi _{2}\rangle }%
+}|{b|^{2})}}  \notag \\
&&{+|b|\left\Vert {a|{{\psi _{1}\rangle }}+b}|{{\psi _{2}\rangle }}%
\right\Vert \big]}  \label{eq9}
\end{eqnarray}
from which it follows that
\begin{equation}
\left\Vert {a|{{\psi _{1}\rangle }}+b}|{{\psi _{2}\rangle
}}\right\Vert E_{g}(|{\Gamma \rangle )}\leq A  \label{eq10}
\end{equation}
where the upper bound $A$ is given in Eq. (\ref{a}). The upper
bound $B$ in Eq. (\ref{b}) is obtained from $A$  by simply
exchanging $a|\psi_1\rangle$ and $b|\psi_2\rangle$. The upper
bound ${\left\Vert {a}|{{\psi _{1}\rangle }+b}|{{\psi _{2}\rangle
}}\right\Vert ^{2}}$ in Eq. (\ref{eq7}) is due to the fact that
the geometric measure is less than or equal to
$1$.$\hfill\blacksquare$

Example 1: Consider the following superposed state
\begin{equation}
|\Gamma \rangle =a|\text{{GHZ}}\rangle +b|W\rangle ,  \label{eq11}
\end{equation}
where $\left\vert \text{{GHZ}}\right\rangle =(1/\sqrt{2})({|{000\rangle }+}|{%
111\rangle )}$ and $\vert W\rangle =(1/\sqrt{3})({|{001\rangle }+}|{{%
010\rangle }+}|{100\rangle )}$. Without loss of generality, we
assume that $a$ and $b$ are both positive real numbers with $a^2 +
b^2 = 1$. The geometric measures of $|\text{{GHZ}}\rangle$ and
$\vert W\rangle$ have been computed in Ref. \cite{Wei:2003} to be
$E_g ( {|\text{{GHZ}}\rangle} ) = 1/2$ and $E_g ( {|W\rangle }) =
5/9$. Inserting these results into Eq. (\ref{eq2}) and Eq.
(\ref{eq7}) yields
\begin{eqnarray*}
E_{g}(|\Gamma \rangle ) &\geq &\max \Big\{{-\frac{1}{18}a^{2}-\frac{4}{3\sqrt{2}}%
a\sqrt{1-a^{2}}+\frac{5}{9},\,0}\Big\},  \\
E_{g}(|\Gamma \rangle ) &\leq &\min \Big\{{\frac{1}{1-a}\left( {\frac{35}{18}%
a^{2}+\frac{3}{2}a+\frac{5}{9}}\right) ,}  \notag \\
&&{\frac{1}{1-\sqrt{1-a^{2}}}\left( {-\frac{37}{18}a^{2}+\frac{13}{9}\sqrt{%
1-a^{2}}+\frac{23}{9}}\right),1 \Big\}.}
\end{eqnarray*}
The lower and upper bounds vs $a$ are shown in Fig.1.

For a superposition of more than two components we can prove the
following proposition by the same way as proving Theorem 1.

\textbf{Proposition:} \emph{For a superposed state} $\left\vert \Gamma \right\rangle =\dfrac{a_{1}\left\vert {\psi _{1}}%
\right\rangle +\cdots +a_{n}\left\vert {\psi _{n}}\right\rangle
}{\left\Vert
{a_{1}\left\vert {\psi _{1}}\right\rangle +\cdots +a_{n}\left\vert {\psi _{n}%
}\right\rangle }\right\Vert }$ \emph{with} $\sum\limits_{i = 1}^n
{\left| {a_i } \right|^2 = 1} $, \emph{the following inequality
holds}

\begin{widetext}
\begin{equation}
\begin{array}{l}
\left\Vert {a_{1}\left\vert {\psi _{1}}\right\rangle +\cdots
+a_{n}\left\vert {\psi _{n}}\right\rangle }\right\Vert ^{2}E_{g}\left( {%
\left\vert \Gamma \right\rangle }\right)  \\
\geq \max \Big\{{\left\vert {a_{1}}\right\vert ^{2}E_{g}\left( {\left\vert {%
\psi _{1}}\right\rangle }\right) +\cdots +\left\vert
{a_{n}}\right\vert ^{2}E_{g}\left( {\left\vert {\psi
_{n}}\right\rangle }\right) +\dsum\limits_{k,l=1,k\neq
l}^{n}{\big[{{a_{k}^{\ast
}a_{l}\langle {\psi _{k}|\psi _{l}\rangle }} -\left\vert {a_{k}a_{l}}%
\right\vert \sqrt{1-E_{g}\left( {\left\vert {\psi _{k}}\right\rangle }%
\right) }\sqrt{1-E_{g}\left( {\left\vert {\psi _{l}}\right\rangle }\right) }}%
\,\big]},0}\Big\}. \\
\end{array}
\label{eq14}
\end{equation}
\end{widetext}

\begin{figure}[ptb]
\includegraphics[scale=0.72,angle=0]{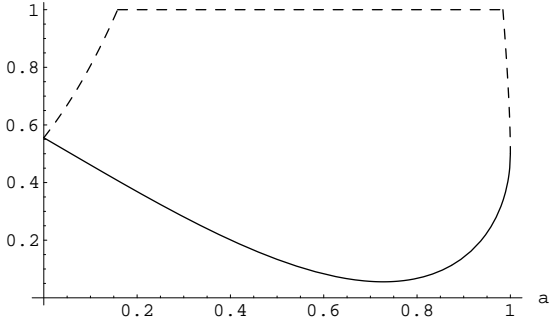}\caption{
The lower and upper bounds of geometric measure of $\left| \Gamma
\right\rangle $ vs $a$. The dash line corresponds to the upper
bound, and the solid line corresponds to the lower bound.}
\label{fig1}%
\end{figure}
In the remainder of this paper, we shall derive an upper bound and a
lower bound in terms of the multipartite q-squashed entanglement
\cite{Christandl:2003}. For an $N$-partite state $\rho _{A_1 ,\ldots
,A_N }$, the q-squashed entanglement is defined as
\begin{equation}
E_{\mathrm{sq}}^{\mathrm{q}}({\rho _{A_{1},\ldots ,A_{N}})}=\inf I({%
A_{1}:A_{2}:\cdots :A_{N}|E)},  \label{eq15}
\end{equation}
where the infimum is taken over all states $\sigma _{A_1 ,\ldots
,A_N ,E} $, that are extensions of $\rho _{A_1 ,\ldots ,A_N } $,
i.e., $\mathrm{Tr}_E \sigma = \rho $. For an $N$-partite pure state
$| \Gamma \rangle _{A_1 ,\ldots, A_N } $, we have\begin{equation}
E_{\mathrm{sq}}^{\mathrm{q}}({\left\vert \Gamma \right\rangle
_{A_{1},\ldots ,A_{N}}})=S({\rho _{A_{1}}})+\cdots +S({\rho
_{A_{N}}}), \label{eq}
\end{equation}
where $\rho _{A_k }$ is obtained from $|\Gamma\rangle\langle\Gamma|$
by tracing out the $k$-th component. We can prove the following
theorem:

\textbf{Theorem 3} \emph{Let} $\left| {\psi _1 }
\right\rangle $ and $\left| {\psi _2 } \right\rangle $ \emph{be
arbitrary normalized N-partite pure states. The q-squashed
entanglement of their superposed state} $\left\vert \Gamma \right\rangle =\dfrac{a\left\vert {\psi _{1}}%
\right\rangle +b\left\vert {\psi _{2}}\right\rangle }{\left\Vert {%
a\left\vert {\psi _{1}}\right\rangle +b\left\vert {\psi _{2}}\right\rangle }%
\right\Vert }$ \emph{with} $\left| a \right|^2 + \left| b \right|^2
= 1$ \emph{satisfies}
\begin{eqnarray}
&&\left\Vert {a\left\vert {\psi _{1}}\right\rangle +b\left\vert {\psi _{2}}%
\right\rangle }\right\Vert ^{2}E_{\mathrm{sq}}^{\mathrm{q}}\left( {%
\left\vert \Gamma \right\rangle }\right)   \notag \\
&\leq &2\Big[ {\left\vert a\right\vert ^{2}E_{\mathrm{sq}}^{\mathrm{q}%
}\left( {\left\vert {\psi _{1}}\right\rangle }\right) +\left\vert
b\right\vert ^{2}E_{\mathrm{sq}}^{\mathrm{q}}\left( {\left\vert {\psi _{2}}%
\right\rangle }\right) +Nh_{2}\big( {\left\vert a\right\vert ^{2}}\big) }%
\Big] .  \notag \\
&&  \label{eq16}
\end{eqnarray}
Proof: To prove this, let us consider the state
\begin{equation}
{\rho }_{A_{1}}^{\prime }=|a|^{2}\mathrm{Tr}_{A_{2},\ldots
,A_{N}}(|{{\psi
_{1}\rangle \langle \psi _{1}|)}}+|b|^{2}\mathrm{Tr}_{A_{2},\ldots ,A_{N}}(|{%
{\psi _{2}\rangle \langle \psi _{2}|)}},  \label{eq17}
\end{equation}
Recalling the property $S({|a|^{2}\rho +}|{b|^{2}\sigma )}\leq
|a|^{2}S(\rho )+|b|^{2}S(\sigma )+h_{2}({|a|^{2})}$, where $h_2 (x)
= - x\log _2 x - (1 - x)\log _2 (1 - x)$ is the binary entropy
function, we have
\begin{eqnarray}
S({{\rho }_{A_{1}}^{\prime })} &\leq
&|a|^{2}S(\mathrm{Tr}{_{A_{2},\ldots
,A_{N}}}|{{\psi _{1}\rangle \langle \psi _{1}|})}  \notag \\
&&+|b|^{2}S(\mathrm{Tr}{_{A_{2},\ldots ,A_{N}}}|{{\psi _{2}\rangle
\langle
\psi _{2}|})}+h_{2}(|{a|^{2}).}  \notag \\
&&  \label{eq18}
\end{eqnarray}
On the other hand, ${\rho }'_{A_1 } $ can also be written as
\begin{eqnarray}
{\rho }_{A_{1}}^{\prime } &=&\frac{\left\Vert {a\left\vert {\psi _{1}}%
\right\rangle +b\left\vert {\psi _{2}}\right\rangle }\right\Vert
^{2}}{2}
\notag \\
&&\times \mathrm{Tr}_{A_{2},\ldots ,A_{N}}\left[ \frac{\left(a\left\vert {\psi _{1}%
}\right\rangle +b\left\vert {\psi _{2}}\right\rangle \right)}{\left\Vert {%
a\left\vert {\psi _{1}}\right\rangle +b\left\vert {\psi _{2}}\right\rangle }%
\right\Vert }\frac{\left(a^{\ast }\left\langle {\psi
_{1}}\right\vert +b^{\ast
}\left\langle {\psi _{2}}\right\vert \right)}{\left\Vert {a\left\vert {\psi _{1}}%
\right\rangle +b\left\vert {\psi _{2}}\right\rangle }\right\Vert
}\right]
\notag \\
&&+\frac{\left\Vert {a\left\vert {\psi _{1}}\right\rangle
-b\left\vert {\psi
_{2}}\right\rangle }\right\Vert ^{2}}{2}  \notag \\
&&\times \mathrm{Tr}_{A_{2},\ldots ,A_{N}}\left[  \frac{\left(a\left\vert {\psi _{1}%
}\right\rangle -b\left\vert {\psi _{2}}\right\rangle \right)}{\left\Vert {%
a\left\vert {\psi _{1}}\right\rangle -b\left\vert {\psi _{2}}\right\rangle }%
\right\Vert }\frac{\left(a^{\ast }\left\langle {\psi
_{1}}\right\vert -b^{\ast
}\left\langle {\psi _{2}}\right\vert \right)}{\left\Vert {a\left\vert {\psi _{1}}%
\right\rangle -b\left\vert {\psi _{2}}\right\rangle }\right\Vert
}\right] .
\notag \\
&&  \label{eq19}
\end{eqnarray}
From the concavity of von Neumann entropy one has
\begin{equation}
\begin{array}[b]{l}
\dfrac{\left\Vert {a\left\vert {\psi _{1}}\right\rangle +b\left\vert
{\psi
_{2}}\right\rangle }\right\Vert ^{2}}{2}S(\mathrm{Tr}{_{A_{2},\ldots ,A_{N}}}%
|{\Gamma \rangle \langle \Gamma |)} \\
+\dfrac{\left\Vert {a\left\vert {\psi _{1}}\right\rangle
-b\left\vert {\psi
_{2}}\right\rangle }\right\Vert ^{2}}{2}S(\mathrm{Tr}{_{A_{2},\ldots ,A_{N}}}%
|{\bar{\Gamma}\rangle \langle \bar{\Gamma}|)}\leq S({{\rho
}_{A_{1}}^{\prime
}),}%
\end{array}
\label{eq20}
\end{equation}
where $\left\vert \bar{\Gamma}\right\rangle =\dfrac{a\left\vert {\psi _{1}}%
\right\rangle -b\left\vert {\psi _{2}}\right\rangle }{\left\Vert {%
a\left\vert {\psi _{1}}\right\rangle -b\left\vert {\psi _{2}}\right\rangle }%
\right\Vert }$. Combining Eq. (\ref{eq18}) and Eq. (\ref{eq20})
leads to
\begin{eqnarray}
&&\left\Vert {a\left\vert {\psi _{1}}\right\rangle +b\left\vert {\psi _{2}}%
\right\rangle }\right\Vert ^{2}S(\mathrm{Tr}{_{A_{2},\ldots
,A_{N}}}|{\Gamma
\rangle \langle \Gamma |)}  \notag \\
&&+\left\Vert {a\left\vert {\psi _{1}}\right\rangle -b\left\vert {\psi _{2}}%
\right\rangle }\right\Vert ^{2}S(\mathrm{Tr}{_{A_{2},\ldots ,A_{N}}}|{\bar{%
\Gamma}\rangle \langle \bar{\Gamma}|)}  \notag \\
&\leq &2\big[|a|^{2}S(\mathrm{Tr}{_{A_{2},\ldots ,A_{N}}}|{\psi
_{1}\rangle
\langle \psi _{1}|)}  \notag \\
&&+|b|^{2}S(\mathrm{Tr}{_{A_{2},\ldots ,A_{N}}}|{\psi _{2}\rangle
\langle \psi _{2}|)+h}_{2}(|a|^{2})\big].  \label{eq21}
\end{eqnarray}
Since $S(\mathrm{Tr}{_{A_{2},\ldots ,A_{N}}}|{\bar{\Gamma}\rangle \langle \bar{%
\Gamma}|)}\geq 0$, it follows that
\begin{eqnarray}
&&\left\Vert {a\left\vert {\psi _{1}}\right\rangle +b\left\vert {\psi _{2}}%
\right\rangle }\right\Vert ^{2}S(\mathrm{Tr}{_{A_{2},\ldots
,A_{N}}}|{\Gamma
\rangle \langle \Gamma |)}  \notag \\
&&\leq 2\big[|a|^{2}S(\mathrm{Tr}{_{A_{2},\ldots ,A_{N}}}|{\psi
_{1}\rangle
\langle \psi _{1}|)}  \notag \\
&&+|b|^{2}S(\mathrm{Tr}{_{A_{2},\ldots ,A_{N}}}|{\psi _{2}\rangle
\langle \psi _{2}|)+h}_{2}(|a|^{2})\big].  \label{eq22}
\end{eqnarray}
Similarly, we can deduce the following inequalities
\begin{eqnarray}
&&\left\Vert {a\left\vert {\psi _{1}}\right\rangle +b\left\vert {\psi _{2}}%
\right\rangle }\right\Vert ^{2}S(\mathrm{Tr}{_{A_{1},\ldots
,A_{k-1}A_{k+1},\ldots ,A_{N}}}|{\Gamma \rangle \langle \Gamma |)}  \notag \\
&\leq &2\big[|a|^{2}S(\mathrm{Tr}{_{A_{1},\ldots
,A_{k-1}A_{k+1},\ldots
,A_{N}}}|{\psi _{1}\rangle \langle \psi _{1}|)}  \notag \\
&&+|b|^{2}S(\mathrm{Tr}{_{A_{1},\ldots ,A_{k-1}A_{k+1},\ldots
,A_{N}}}|{\psi _{2}\rangle \langle \psi _{2}|)+h}_{2}(|a|^{2})\big]\notag \\
\label{eq23}
\end{eqnarray}
for $k = 1,\ldots ,N$. Adding all these inequalities together and
using Eq. (\ref{eq}), the advertised inequality in Eq. (\ref{eq16})
is proved.$\hfill\blacksquare$

Example 2: Consider the following $N$-partite states:
\begin{eqnarray}
|{\psi _{1}\rangle } &=&\sqrt{\frac{1}{10}}|1\rangle ^{\otimes N}+\sqrt{%
\frac{9}{10}}\sqrt{\frac{1}{d-1}}\big(|2\rangle ^{\otimes N}{+\cdots
+}|d\rangle
^{\otimes N}\big),  \notag \\
|{\psi _{2}\rangle } &=&\sqrt{\frac{1}{10}}|1\rangle ^{\otimes N}-\sqrt{%
\frac{9}{10}}\sqrt{\frac{1}{d-1}}\big(|2\rangle ^{\otimes N}{+\cdots
+}|d\rangle
^{\otimes N}\big),  \notag \\
a  &=&-b =\frac{1}{\sqrt{2}}.  \label{eq25}
\end{eqnarray}
We fix $d = 11$ and consider $N \leq 8$. For each superposed state
we calculate $E_{\mathrm{sq}}^{\mathrm{q}}(|\Gamma\rangle) $ and
its upper bound. The results are shown in Fig.2(a). On the other
hand, for $N = 3$ and $d\le 8$, the values of
$E_{\mathrm{sq}}^{\mathrm{q}}(|\Gamma\rangle) $ and the
corresponding upper bounds are shown in Fig.2(b). One sees that
each q-squashed entanglement diverges from its upper bound not too
much in all these cases.

Recently, Gour \cite{Gour:2007} derived tight lower and upper bounds
on the entanglement (von Neumann entropy) of a superposition of two
bipartite states in terms of the entanglement of the two states
constituting the superposition. The new upper bound is tighter than
the one presented in \cite{Linden:2006}. Gour's upper bound leads
immediately to a new upper bound for the q-squashed entanglement of
an $N$-partite pure state; the new upper bound is more stringent
than the one given in Theorem 3.

\textbf{Theorem 4.} \emph{Let} $\left| {\psi _1 } \right\rangle $
\emph{and} $\left| {\psi _2 } \right\rangle $ \emph{be arbitrary
normalized N-partite pure states. The q-squashed entanglement of
their superposed states} $\left\vert \Gamma \right\rangle =\dfrac{a\left\vert {\psi _{1}}%
\right\rangle +b\left\vert {\psi _{2}}\right\rangle }{\left\Vert {%
a\left\vert {\psi _{1}}\right\rangle +b\left\vert {\psi _{2}}\right\rangle }%
\right\Vert }$ \emph{with} $\left| a \right|^2 + \left| b \right|^2
= 1$ \emph{satisfies}
\begin{equation}
\left\Vert {a\left\vert {\psi _{1}}\right\rangle +b\left\vert {\psi _{2}}%
\right\rangle }\right\Vert ^{2}E_{\mathrm{sq}}^{\mathrm{q}}\left( {%
\left\vert \Gamma \right\rangle }\right) \leq f\left( t\right)
\label{eq24}
\end{equation}
\emph{ for all} $0 \le t \le 1$, \emph{where}
\begin{eqnarray*}
f\left( t\right)  &=&\frac{t|b|^{2}+({1-t)}|a|^{2}}{t({1-t)}}\big[{tE_{%
\mathrm{sq}}^{\mathrm{q}}}(|{{{\psi _{1}\rangle )}}} \\
&&{+}({{1-t)}E_{\mathrm{sq}}^{\mathrm{q}}(|{{{\psi _{2}\rangle )}}}+Nh_{2}}({%
t)\big].}
\end{eqnarray*}
Here, the minimum of the function $f(t)$ is achieved when $t$
satisfies the equation
\begin{equation*}
\frac{|a|^{2}({1-{t)}}^{2}}{|b|^{2}{t}^{2}}=\frac{E_{\mathrm{sq}}^{\mathrm{q}%
}(|{{\psi _{1}\rangle )}}-N\log_2
{t}}{E_{\mathrm{sq}}^{\mathrm{q}}(|{{\psi _{2}\rangle )}}-N\log
_2({1-{t})}}
\end{equation*}

\begin{figure}[ptb]
\includegraphics[scale=0.9,angle=0]{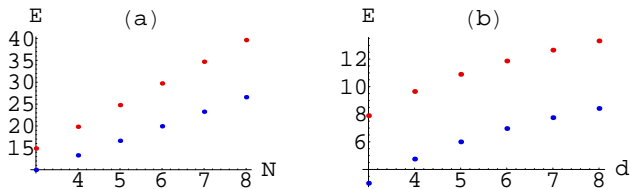}\caption{Color online. (a)Plots of the q-squashed entanglement of
superposed states (blue dots) and upper bounds (red dots) for $d =
11$ and $N \le 8$,(b)$N = 3$ and $d \le 8$.}
\label{fig2}%
\end{figure}

Gour's lower bound, on the other hand, enables us to obtain a lower
bound for the q-squashed entanglement of an $N$-partite pure state:

\textbf{Theorem 5}. \emph{Let} $\left| {\psi _1 } \right\rangle $
\emph{and} $\left| {\psi _2 } \right\rangle $ \emph{be arbitrary
N-partite pure states, and let} $\left| \Gamma \right\rangle =
a\left| {\psi _1 } \right\rangle + b\left| {\psi _2 }
\right\rangle $ \emph{be a normalized state. Then the q-squashed
entanglement of} $\left| \Gamma \right\rangle $ \emph{satisfies}
\begin{equation}
E_{\mathrm{sq}}^{\mathrm{q}}\left( {\left\vert \Gamma \right\rangle
}\right) \geq \max \left\{ {C\left( t\right) ,D\left( t\right)
}\right\} . \label{eq26}
\end{equation}
\emph{for all} $0 \le t \le 1$, \emph{where}
\begin{equation*}
\begin{array}{l}
C\left( t\right) =\frac{\left( {1-t}\right) \left\vert b\right\vert ^{2}}{%
1-t\left( {1-\left\vert a\right\vert ^{2}}\right) }E_{\mathrm{sq}}^{\mathrm{q%
}}\left( {\left\vert {\psi _{2}}\right\rangle }\right) -\frac{1-t}{t}E_{%
\mathrm{sq}}^{\mathrm{q}}\left( {\left\vert {\psi _{1}}\right\rangle }%
\right) -\frac{N}{t}h_{2}\left( t\right) , \\
D\left( t\right) =\frac{\left( {1-t}\right) \left\vert a\right\vert ^{2}}{%
1-t\left( {1-\left\vert b\right\vert ^{2}}\right) }E_{\mathrm{sq}}^{\mathrm{q%
}}\left( {\left\vert {\psi _{1}}\right\rangle }\right) -\frac{1-t}{t}E_{%
\mathrm{sq}}^{\mathrm{q}}\left( {\left\vert {\psi _{2}}\right\rangle }%
\right) -\frac{N}{t}h_{2}\left( t\right) .%
\end{array}%
\end{equation*}

The maximum of $C\left( t \right)$ is obtained when
\begin{equation*}
\frac{\left\vert a\right\vert ^{2}\left\vert b\right\vert ^{2}{t}^{2}}{[{1-}(%
{{1-\left\vert a\right\vert
^{2})t]}}^{2}}E_{\mathrm{sq}}^{\mathrm{q}}\left(
{\left\vert {\psi _{2}}\right\rangle }\right) =E_{\mathrm{sq}}^{\mathrm{q}%
}\left( {\left\vert {\psi _{1}}\right\rangle }\right) -N\log_2 (
{1-t}) .
\end{equation*}
The analogous formula applies to $D\left( t \right)$.

Analogous to Ref. \cite{Linden:2006}, we can show that if the
entanglement is quantified by the multipartite q-squashed
entanglement, then two states of high fidelity to one another do
not necessarily have nearly the same entanglement.

Example 3: Suppose $\left| {\psi _1 } \right\rangle = \left| {000}
\right\rangle $, and $\left\vert {\psi _{2}}\right\rangle
=\sqrt{1-\varepsilon }\left\vert {\psi _{1}}\right\rangle
+\sqrt{\varepsilon /d}(|111\rangle +\cdots +|ddd\rangle)$. It is
easy to show that $E_{\mathrm{sq}}^{\mathrm{q}}\left( {\left\vert {\psi _{1}}\right\rangle }%
\right) =0$ and $E_{\mathrm{sq}}^{\mathrm{q}}\left( {\left\vert {\psi _{2}}%
\right\rangle }\right) =3\left[ {-\left( {1-\varepsilon }\right)
\log
_{2}\left( {1-\varepsilon }\right) -d\left( {\frac{\varepsilon }{d}\log _{2}%
\frac{\varepsilon }{d}}\right) }\right] \approx 3\varepsilon \log
_{2}d$. The fidelity $\left| {\left\langle {\psi _1 } \right|\left.
{\psi _2 } \right\rangle } \right|^2 = 1 - \varepsilon $ approaches
one for small $\varepsilon $, while the difference in the q-squashed
entanglement of $\left| {\psi _1 } \right\rangle $ and $\left| {\psi
_2 } \right\rangle $ can be as large as we expect if we choose an
appropriate $d$.

Summarizing, we have presented lower and upper bounds on the
entanglement of the multipartite superposition state in terms of the
geometric measure and q-squashed entanglement measure, respectively.
Our results partly solve the open problem proposed in Ref.
\cite{Linden:2006}. In view of the fact that the geometric measure
and the q-squashed entanglement measure are both multipartite
entanglement measure, our results may find useful applications in
future manipulations of multipartite entanglement.

We thank D. Yang for valuable suggestions and K. Chen for bringing
Ref. \cite{Niset:2007} and \cite{Cavalcanti:2007} to our attention.
This work is supported by the NNSF of China, the CAS, the National
Fundamental Research Program (Grant No. 2006CB921900), and Anhui
Provincial Natural Science Foundation (Grant No. 070412050).

\emph{Note added.} After completing this manuscript, We became aware
of two recently related papers by J. Niset \emph{et al.}
\cite{Niset:2007} and D. Cavalcanti \emph{et al.}
\cite{Cavalcanti:2007}.

\end{document}